\pdfoutput=1
\documentclass{aastex61}
\usepackage{subfloat}
\usepackage{graphicx}
\begin{document}
\title{Galaxy Ellipticity Measurements in the Near-Infrared for Weak Lensing}
\correspondingauthor{Bomee Lee}
\email{bomee@ipac.caltech.edu}

\author{Bomee Lee}
\affil{MS314-6, Infrared Processing and Analysis Center, California Institute of Technology, Pasadena, CA 91125, USA}
\author{Ranga-Ram Chary}
\affil{MS314-6, Infrared Processing and Analysis Center, California Institute of Technology, Pasadena, CA 91125, USA}
\author{Edward L. Wright}
\affil{UCLA Astronomy, P.O. Box 951547, Los Angeles, CA 90095, USA}

\begin{abstract}

We investigate the value of the near-infrared imaging from upcoming surveys for constraining the ellipticities of galaxies. 
We select galaxies between $0.5 \le z < 3$ that are brighter than expected {\it Euclid} sensitivity limits from the GOODS-S and N fields 
in CANDELS. The co-added CANDELS/HST V+I and J+H images are degraded in resolution and sensitivity to simulate {\it Euclid}-quality 
optical and near-infrared (NIR) images. We then run GALFIT on these simulated images and find that optical and NIR 
provide similar performance in measuring galaxy ellipticities at redshifts $0.5 \le z < 3$. 
At $z>1.0$, the NIR-selected source density is higher by a factor of 1.4 and therefore the standard error in NIR-derived 
ellipticities is about 30\% smaller, implying a more precise ellipticity measurement. 
The good performance of the NIR is mainly because galaxies have an intrinsically smoother light distribution in the NIR bands than in the optical, the 
latter tracing the clumpy star-forming regions. In addition, the NIR bands have a higher surface brightness 
per pixel than the optical images, while being less affected by dust attenuation. 
Despite the worse spatial sampling and resolution of {\it Euclid} NIR compared to optical, the NIR approach 
yields equivalent or more precise galaxy ellipticity measurements. If systematics that affect shape such as dithering strategy and 
point spread function undersampling can be mitigated, inclusion of the NIR can improve galaxy ellipticity measurements over all redshifts. 
This is particularly important for upcoming weak lensing surveys, such as with {\it Euclid} and WFIRST.
 \end{abstract}

\section{Introduction}

Weak gravitational lensing (WL) is a slight deflection of light rays from distant galaxies when they propagate through the tidal gravitational 
field of intervening large scale structure. The amplitude of the WL distortion can be used to map dark matter and measure dark energy by statistically 
quantifying the shear distortions encoded in the observed shapes of background galaxies, namely galaxy ellipticities 
\cite[e.g.][]{kai95, bar01}. The ellipticities of galaxies are typically distorted only about a per cent by WL \citep{tro15}, so the WL signal in individual galaxies 
is challenging to detect. WL measurements thus rely on averaging over a very large sample to obtain 
the distortions and sufficiently unbiased estimates of galaxy shapes, which in turn require a correction for the 
impact of the point spread function (PSF) of the telescope. In that sense, WL observations demand high quality images because it requires a large number 
density of resolved galaxies and high signal-to-noise ratio (SNR), while minimizing the PSF corrections and related systematic uncertainties, with well-sampled PSFs 
\citep{mas13, sch18}. 

{\it Euclid} \citep{lau11} is a survey mission designed to understand the expansion and growth history of the Universe, and is scheduled to launch in the next decade.
{\it Euclid} will image 15,000\,deg$^{2}$ of sky 
in one broad optical band VIS spanning 550 to 920\,nm, and three additional near-infrared (NIR) bands ($Y$, $J$, and $H$).
{\it Euclid} will detect cosmic shear with VIS by measuring ellipticities of 
$\sim30$ resolved galaxies per arcmin$^2$ with a resolution better than 0.18$\arcsec$ (PSF FWHM) with 0.1$\arcsec$ pixels. 
The near-infrared bands will primarily be used to derive photometric 
redshifts for the weak lensing sample, in conjunction with ground-based observations at visible wavelengths. The {\it Euclid} wide 
survey is expected to provide WL galaxy shape measurements for 1.5 billion galaxies with space-quality resolution.

To measure WL through surveys, one should measure galaxy ellipticities and its uncertainty, including systematics, very accurately. 
In particular, it is necessary to measure the shapes of typically faint and small, distant galaxies with high-SNR observations. In this work, 
we demonstrate that NIR bands result in a comparable or more precise galaxy ellipticity measurement compared to optical bands for WL studies 
despite their worse spatial resolution (0.3$\arcsec$ compared to 0.18$\arcsec$) and pixel sampling (0.3$\arcsec$ vs. 0.1$\arcsec$ pixel scale). 
There are several advantages to using NIR bands \citep{tun17}; first, NIR wavelengths sample the rest-frame optical light, 
which traces the older stellar population (hence the bulk of stellar mass) and is less affected by dust extinction. The VIS band covers 
the rest-frame UV and blue wavelengths, which predominantly traces emission from star-forming regions \citep{dic00}. In particular, the shapes of 
galaxies as seen in the rest-frame UV are more clumped and irregularly distributed than older stellar populations. The second 
advantage is that galaxies in the NIR bands have an intrinsically smoother light distribution resulting in a lower shape noise than in the optical \citep{sch18}. 
Third, NIR images of galaxies have a higher surface brightness with more than nine times the number of source photons per pixel, based on 
a calculation using images in this study; this is at least
partly due to the relative importance of the bulge compared to the disk as a function of wavelength. 
Finally, we find that the NIR bands are sensitive to a larger number density of distant galaxies than the VIS band (see Section 2).

In this paper, we study the shapes of the galaxy sample expected from {\it Euclid}-quality imaging and forecast how we can 
improve the shape measurement by using co-added NIR images\footnote{The simulated {\it Euclid} images in this paper do not have {\it Euclid}-specific systematics
dealing with dither strategy, field distortion, PSF variations and intrapixel quantum efficiency variations which will be investigated in the future. However, it should be noted that these affect
both optical and near-infrared images.}. To do that, we select galaxies from 
{\it HST}/CANDELS (Cosmic Assembly Near-infrared Extragalactic Legacy Survey; \cite{gro11}; \cite{koe11}) 
observations satisfying the {\it Euclid} sensitivity limits 
and simulate {\it Euclid}-resolution images. The structure of this paper as follows. The sample selection using CANDELS data
 is introduced in Section 2. We describe the procedure of 
simulating {\it Euclid}-quality optical and NIR images from {\it HST} images in Section 3. In Section 4, we explain how GALFIT (Peng et al. 2010) is used to measure the ellipticity 
of galaxies after accounting for the PSF, and compare the ellipticities obtained from GALFIT in simulated-{\it Euclid} and CANDELS images. Finally, we present
our conclusions in Section 5.

\section{Initial Sample Selection}

We select a sample of galaxies at optical and near-infrared (NIR) wavelengths from the HST/CANDELS survey that closely resembles 
the {\it Euclid} weak lensing (WL) sample. Among the five CANDELS fields, we use the GOODS-S and 
GOODS-N fields which include the CANDELS Deep survey and covers about 340 $arcmin^{2}$ in V (0.606$\rm \mu m$), I (0.814$\rm \mu m$), 
J (1.25$\rm \mu m$), and H (1.6$\rm \mu m$). These fields are several magnitudes deeper than the {\it Euclid} survey.
We use CANDELS photometric redshifts measured for all galaxies by \cite{dah13}, 
unless spectroscopic redshifts are available. For the WL shape measurement, the {\it Euclid} survey 
will detect galaxies in a broad optical R+i+z band (VIS: 0.55--0.92$\rm \mu m$) down to 24.5 mag (10 $\sigma$). It will use three additional NIR bands 
(Y, J, H in the range of 0.92--2.0 $\mu m$) reaching AB mag 24 (5 $\sigma$) in each. To achieve the required dark energy figure of merit through weak lensing, the surface density 
of resolved galaxies needs to be at least 30 $arcmin^{-2}$ \citep[Euclid Red book;][]{lau11}. 

\begin{figure}
\centering
\includegraphics[width=4.5in]{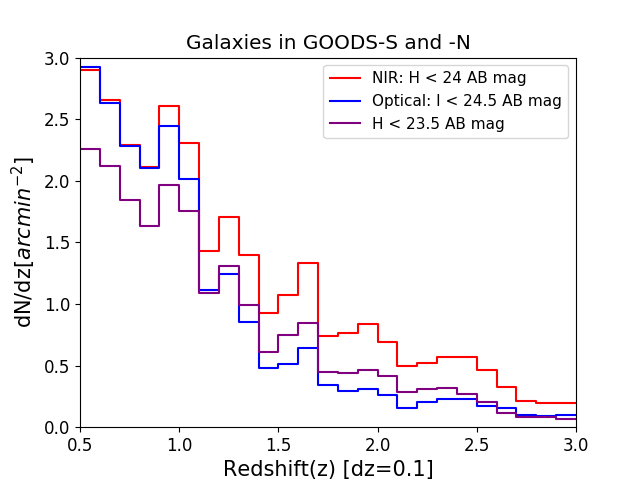}
\caption{Number density of galaxies at a redshift range, $0.5\le z < 3.0$, in the GOODS-S and GOODS-N fields. We compare number density of galaxies 
in redshift bins of width $\delta z=0.1$ of different galaxy samples, selected using expected magnitude depths of the {\it Euclid} survey. 
The red histogram represents a NIR selection of $H< 24$\,mag while the blue histogram is for 
${I} < 24.5$\,mag. We also sub-select galaxies having $H<23.5$\,mag based on the quality of their ellipticity fits and 
present their redshift distribution with the purple histogram (see a further explanation about this selection in Section 4.3). 
At $z>1.0$, the NIR yields a higher surface density of galaxies than in the optical. The median (mean) redshift is  0.95(1.1), 1.1(1.3), and 1.0(1.2) for the 
optical, NIR, and $H<23.5$ selected sample, respectively, within a redshift range of $0.5\le z < 3.0$. }
\label{fig:histz}
\end{figure}

We start by replicating the {\it Euclid} expected sensitivity selection on the CANDELS catalogs. We find that
I $< $24.5 AB mag results in about 30 galaxies per $arcmin^{2}$ with a mean redshift of $\sim 0.9$, which is consistent with the {\it Euclid} requirement. 
Applying the {\it Euclid} $H<24$ mag selection on the CANDELS NIR sample, results in a mean $z\sim1.1$ with about 37 galaxies per $arcmin^{2}$. 
The redshift distribution of each sample selection is shown in Figure~\ref{fig:histz}. 
One clear advantage of the NIR is at $z>1$, where the NIR bands select many more galaxies than the optical. This suppresses the shape noise induced by 
the intrinsic ellipticities of distant galaxies 
if the individual ellipticity uncertainties were similar to that in the optical; we assess the veracity of this in the following sections.
In this study, we specifically use galaxies at a redshift range of $0.5\le z < 3$ to compare ellipticities estimated from optical and NIR images.

\section{{\it Euclid} images made from CANDELS/{\it HST} images}

\begin{figure}
\centering
\includegraphics[width=7in]{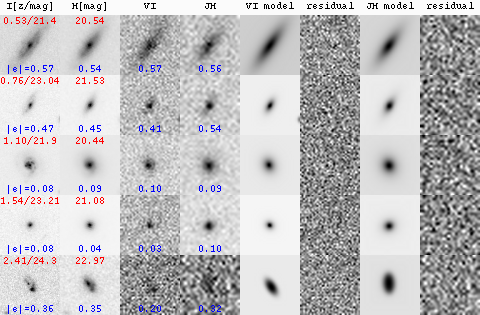}
\caption{Postage stamps of five galaxies in the GOODS-S field at different redshifts. Each image covers an area of 4$\arcsec\times 4\arcsec$. 
From left to right, CANDELS I, H, simulated {\it Euclid} V+I, and J+H images, GALFIT model fit to the Euclid VI image and residual, model fit to the {\it Euclid}
JH image and residual. Redshift, AB magnitude of I and H bands for each galaxy are given in the top (red texts) and the modulus of the galaxy ellipticity calculated from 
the GALFIT results using respective images are given in the bottom (blue texts). }
\label{fig:post}
\end{figure}

We simulate {\it Euclid} VIS and NIR images using CANDELS/{\it HST} V (0.606$\rm \mu m$), I (0.814$\rm \mu m$), J (1.25$\rm \mu m$), and H (1.6$\rm \mu m$). 
Each of the {\it Euclid} VIS, Y, J and H bands will have four images taken per unit area of sky with 0.1$\arcsec$ pixel scale in the VIS and 0.3$\arcsec$ pixel scale in the NIR
bands \citep{lau11}. For the VIS images, we combine the CANDELS V and I band images which span the bandwidth of the {\it Euclid} VIS band, 0.55$\rm \mu m$--0.92$\rm \mu m$. 
In both the VIS and NIR bands, there will be significant correlated noise if coadded images are made on finer pixel scale with just four frames. 
By combining the $J$ and $H$ band images, we can
both increase the signal to noise in the NIR and drizzle on a factor of two oversampled pixel scale with a point-kernel 
(e.g. CANDELS WIDE survey; \cite{gro11}, HST/ACS COSMOS; \cite{rho07}), thereby minimizing the impact of correlated noise.
We therefore combine the CANDELS $J$ and $H$ band images ({\it Euclid} $J$: 1.16--1.58$\rm \mu m$, $H$: 1.52--2.04$\rm \mu m$). We estimate that 
the {\it Euclid} survey strategy results in a median of between 10 and 11 valid frames per pixel when combining all three bands $Y$, $J$ and $H$.
However, the PSF undersampling in the shortest wavelength band and color gradients across such a wide wavelength range
may introduce other systematics. The impact on the undersampled
PSF as a result of the drizzling and the {\it Euclid} dither strategy is beyond the scope of this work and is currently being investigated. 
Furthermore, since the CANDELS $Y$ band imaging does not cover the entire GOODS-S and -N fields, we avoid including the $Y-$band in this analysis.

The step-by-step procedure to simulate the {\it Euclid}-quality images is outlined below:

\begin{enumerate}
\item Produce cutouts of science and noise images ($rms$) for each galaxy from the large {\it HST}/CANDELS V, I, J, and H mosaics. 

\item Combine V and I or J and H by weighting each pixel according to the weight map (inverse variance), i.e. 
$f_{comb} = (f_1w_1+f_2w_2)/(w_1+w_2)$, where $f_{1,2}$ and $w_{1,2}$ are the pixel values of science image and weight map, 
and $w_{1,2} \sim 1/rms_{1,2}^2 $. Because each pixel value has noise associated with it, and the noise is somewhat heterogeneous 
due to the observing strategy, an inverse variance weighting is 
the optimal approach to combine images and increase the signal-to-noise of co-added images \citep{koe11}.

\item Re-bin the pixel scale from 0.06$\arcsec$ of CANDELS to 0.1$\arcsec$ (V+I) or 0.15$\arcsec$ (J+H). 
By using co-added J+H images, which will double the number of images, the co-added NIR 
images can be drizzled onto a 0.15$\arcsec$ pixel scale, half of the original NIR pixel scale.
This is challenging to do for VIS, since only four frames will be taken. 

\item Smooth the combined images with a Gaussian kernel to correct for the difference in PSF FWHM between {\it HST} and
 {\it Euclid} (for optical, 0.1$\arcsec$ vs. 0.18$\arcsec$; for NIR, 0.18$\arcsec$ vs. 0.3$\arcsec$). 

\item Make noise maps for the V+I and J+H following a random Gaussian distribution with 1$\sigma$ measured from the quoted sensitivity of {\it Euclid} VIS and NIR 
images. For the VIS images, the sensitivity is 24.5 mag at 10 $\sigma$ (estimated from an extended source with a 0.3$\arcsec$ radius; \cite{cro12}). For the NIR
images, the sensitivity is 24 mag at 5 $\sigma$ in each band (measured from a 
point source), respectively. By combining the $J$ and $H$ bands, the effective sensitivity is therefore 24 AB mag at 7$\sigma$.

\item Add the noise maps to the images to obtain the simulated {\it Euclid} VI(V+I) and JH(J+H) images. Since CANDELS' background noise is negligible 
(more than 50 times smaller than that of {\it Euclid}), we do not remove the noise in the CANDELS images before adding {\it Euclid} noise maps.
\end{enumerate}

A few galaxies in the sample (1\% and 3\% in JH and VI, respectively) are not observed in J and V bands because the CANDELS coverage of the field at 
different bands varies slightly. Thus, after excluding these sources, we have 7,248 galaxies for VI and 9,887 galaxies for JH. For illustrative purposes, the 
images of 5 galaxies in the CANDELS I and H bands, the simulated {\it Euclid} VI and JH images, and their GALFIT fits are shown in Figure~\ref{fig:post}. 

\section{Ellipticity measurements using GALFIT}

\subsection{Masking sources}

Weak lensing measurements, due to the small signal, typically rely on averages over a large number of galaxies. 
As a result, they usually require aggressive masks of samples 
to correct systematic effects. In particular, due to the sensitivity limit and spatial resolution, we find that {\it Euclid} will suffer from blending of 
galaxies with nearby objects and non-detections which the higher spatial resolution of the VIS band may be able to reveal. 
As demonstrated in Figure~\ref{fig:post}, the spatial resolution and SNR of simulated Euclid images (3rd and 
4th columns) unsurprisingly appear to be significantly worse than CANDELS images (1st and 2nd). We therefore run 
a source detection algorithm, SExtractor \citep{ber96}, on the 
VI and JH cutouts in Section 3 and remove sources for which the photometry as measured by the SEXtractor, AUTO MAG, deviates from the expected magnitude 
(I or H band magnitude from CANDELS photometry catalog) by more than two times the uncertainty in the difference between derived and expected magnitudes. 
In addition, we exclude galaxies from the original sample which are now offset by more than 0.7$\arcsec$ (VI) and 0.75$\arcsec$ (JH) relative 
to the original positions because it implies that the detection in the {\it Euclid} simulated image is either noise or affected by source confusion. 
After masking out about 13.3\% and 11.4\% sources from Section 3, we have 6,283 and 8,762 galaxies
for VI and JH respectively. At the expected sensitivity limit, we find that we are about 80\% complete at $24.5$\,AB mag in VI and $24.0$\,AB mag in JH.

\subsection{GALFIT}

We measure the ellipticities for the galaxies in the sample using GALFIT \citep{pen02}. GALFIT fits a S\'{e}rsic law to the surface brightness 
profile measured within elliptical isophotes of a galaxy. Importantly, GALFIT includes the PSF in the fitting process. 
The accounting for the PSF is crucial in WL because 
the smoothing from the PSF make galaxies appear rounder than they actually are \citep{hol09} and significantly 
biases the ellipticity measurements. The appropriate PSF model is essential for the accuracy of 
the ellipticity estimation. It is inappropriate to derive a PSF for galaxies from the stars
because stars typically have a Rayleigh-Jeans spectrum across the bandpass while galaxies are significant redder, implying a broader intrinsic PSF.
We therefore construct a model PSF for the CANDELS using the TinyTim software package \citep{kri95} for 
the ACS I band and WFC3 H band by assuming a flat galaxy spectrum ($f_{\nu} \sim constant$). They are 
then re-sampled to the CANDELS pixels scale, 0.06$\arcsec$. For {\it Euclid}, we re-sample I and H Tinytim PSF to the {\it Euclid} pixel scales, 0.1$\arcsec$ (VI) 
and 0.15$\arcsec$ (JH) and, subsequently, smooth with a Gaussian smoothing kernel to correct for the difference in 
PSF FWHM between CANDELS and {\it Euclid}. 

\begin{table}[]
\caption{Number of galaxies per 340 $arcmin^{2}$ (and number of galaxies per $arcmin^{2}$) in five redshift bins between $0.5\le z <3.0$. }
\label{table2}
\scalebox{0.8}{
\begin{tabular}{|p{4cm}p{2cm}ccccc|}
\hline
 & {\begin{tabular}[c]{c}Total\\ (N/arcmin$^2$)\end{tabular}} & $0.5\le z<0.7$ & $0.7\le z<1.0$ & $1.0\le z<1.3$ & $1.3\le z<1.9$ & $1.9\le z<3$ \\ \hline
VI (I\textless24.5 \& R$_e$ \textgreater 0.1") & \begin{tabular}[c]{c}4634\\ (13.6)\end{tabular} & \begin{tabular}[c]{l}1318\\ (3.9)\end{tabular} & \begin{tabular}[c]{l}1571\\ (4.6)\end{tabular} & \begin{tabular}[c]{l}899\\ (2.6)\end{tabular} & \begin{tabular}[c]{l}529\\ (1.6)\end{tabular} & \begin{tabular}[c]{l}317\\ (0.9)\end{tabular} \\ \hline

JH (H\textless23.5 \& R$_{e}$ \textgreater 0.15") & \begin{tabular}[c]{c}4770\\ (14.0)\end{tabular} & \begin{tabular}[c]{l}1042\\ (3.1)\end{tabular} & \begin{tabular}[c]{l}1352\\ (4.0)\end{tabular} & \begin{tabular}[c]{l}1004\\ (2.9)\end{tabular} & \begin{tabular}[c]{l}903\\ (2.6)\end{tabular} & \begin{tabular}[c]{l}469\\ (1.4)\end{tabular} \\ \hline
\end{tabular}}
\end{table}

We let GALFIT fit the images with central position, magnitude, half-light radius (R$_{e}$) 
measured along the major axis, S\'{e}rsic index, axis ratio (q = semi-minor axis/semi-major axis), and 
position angle as free parameters. The SExtractor measurements are used to feed GALFIT with initial guesses for these parameters. 
In each image cutout, neighboring objects detected from the SExtractor are fit simultaneously or masked out if they are less than 2 magnitudes 
fainter than the target galaxy. Any fit resulting in problems (i.e., axis ratio errors $>$1.0) or non-existent results (fits crashed) are excluded. 
According to experiments undertaken by \cite{vdw12}, about 60\% of galaxies have a good fit (GALFIT flag $=$ 0) from 
CANDELS GOODS-S GALFIT catalog, which can be used as reliable measures of ellipticity. We find a slightly higher percentage, 
4,737 galaxies (65.4\%) for VI and 6,449 galaxies (65.2\%) for JH, after excluding all problematic galaxies as discussed in Section 4.1 and bad fits. 
Figure~\ref{fig:post} illustrates best-fit GALFIT model images and residuals images showing 
the difference between model and original image which are dominated by noise. The absolute value of the ellipticity (see the definition in Section 4.3) computed using 
the GALFIT results for those galaxies are given in the bottom of I, H, VI, and JH images. 

Although GALFIT is one of the most popular fitting tools for measuring galaxy shapes, we lose a large amount of our sample due to unreliable fits. 
Furthermore, it is known that the GALFIT is not suitable to fit small, faint galaxies, mainly high redshift galaxies \citep{vdw12, sif15}. 
This is mostly due to the number of parameters that GALFIT tries to fit for, which results in unreliable fits in the low signal to noise ratio regime \citep{Jee}.
The widely used 
shear measurement algorithms uniquely developed for WL, such as the KSB (Kaiser, Squires \& Broadhurst) algorithm 
\citep{kai95, hoe98} and $\it{lens}$fit \citep{mil07, kit08}, might provide better performance in measuring the 
ellipticity of galaxies. However, \cite{sif15} compared the KSB results for bright cluster galaxies to GALFIT shapes and 
showed that the ellipticities measured by both methods are generally consistent. A detailed assessment of the accuracy of ellipticity 
measurements from different techniques is beyond of the scope of this paper and we use GALFIT for our main goal of comparing 
the ellipticities estimated from CANDELS to {\it Euclid}-quality images.

\begin{figure*}
\centering
\epsscale{1.1}
\plottwo{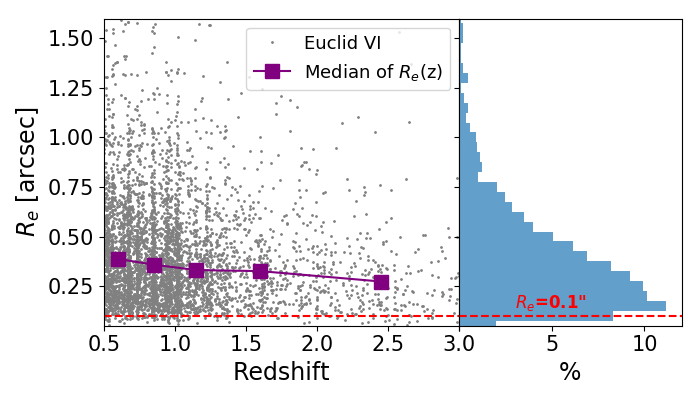}{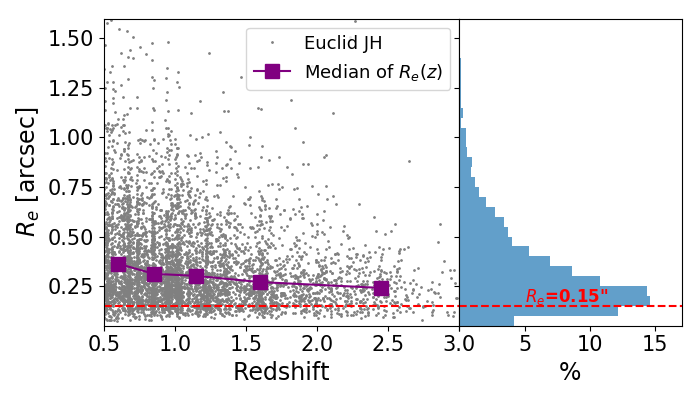}
\caption{Redshift vs. Half-light radius ($R_{e}$ [arcsec]) estimated using GALFIT with VI (left) and JH (right) images with a histogram 
of $R_{e}$ for $I <24.5$\,mag and $H < 23.5$\,mag sample. The median of $R_{e}$ at five redshift bins (in Table 1) are 
overplotted with purple squares. The cuts for the galaxy size are  $R_{e} > 0.1"$ in VI and $R_{e} > 0.15"$ in JH (red dashed lines). 
Using the size cut, we exclude 2.2\% galaxies among $I < 24.5$\,mag sample and 9.8\% galaxies among $H<23.5$\,mag sample.} 
\label{fig:re}
\end{figure*}

\subsection{Comparison of ellipticities between CANDELS and Euclid-quality images}

Typically, only galaxies with a size comparable or larger than the PSF have a well measured shape; so the shape of the smallest galaxies 
becomes ill-defined. Also, as the SNR decreases, the ability to measure galaxy shapes decreases since the imaging data is only sensitive to the highest
surface brightness regions of the galaxy. Therefore, the galaxy samples for shape 
measurement require a lower limit to the SNR of about 10, and the radius of the galaxy larger than 1.25 times the PSF FWHM (Euclid Red book: \citet{lau11}). 
In order to satisfy those requirements, we restrict the galaxy sample in the NIR band to have H $< 23.5$\,mag ($\sim$ SNR $>11$ for J+H), 
and a half-light radius ($R_{e}$) measured from GALFIT on the simulated {\it Euclid} JH image of larger than half of the NIR PSF FWHM of {\it Euclid} (R$_{e} > $ 0.15$\arcsec$). 
For the optical, we select galaxies having I $< 24.5$\,mag and SNR(I) $>10$ with a size limit, $R_{e}$ measured from VI images $> 0.1\arcsec$. 
This is about half the FWHM of the PSF in the VIS band. In Figure~\ref{fig:re},
we show the distribution of $R_{e}$ as a function of a redshift 
for the optical and NIR sample with $I<24.5$\,mag (left) and $H<23.5$\,mag (right), respectively. 
About 2.2\% of optical and 9.8\% of NIR selected galaxies have $R_{e} < 0.1\arcsec$ and $R_{e} < 0.15\arcsec$, respectively. As a final sample for analyzing ellipticities, 
we use 4,634 and 4,770 galaxies for the optical and NIR respectively. In Table 2, the number densities of the galaxy samples in the optical and NIR 
are listed at five different redshift bins. 
The relatively high SNR cut of H$< 23.5$\,mag results in a similar total number 
density of galaxies with the VI band as also shown in Figure~\ref{fig:histz}, but still translates to a higher number density by a factor of 1.4 at $z >1$ (6.9 vs. 5.1 arcmin$^{-2}$). 

\begin{figure*}
\centering
\epsscale{1.1}
\plotone{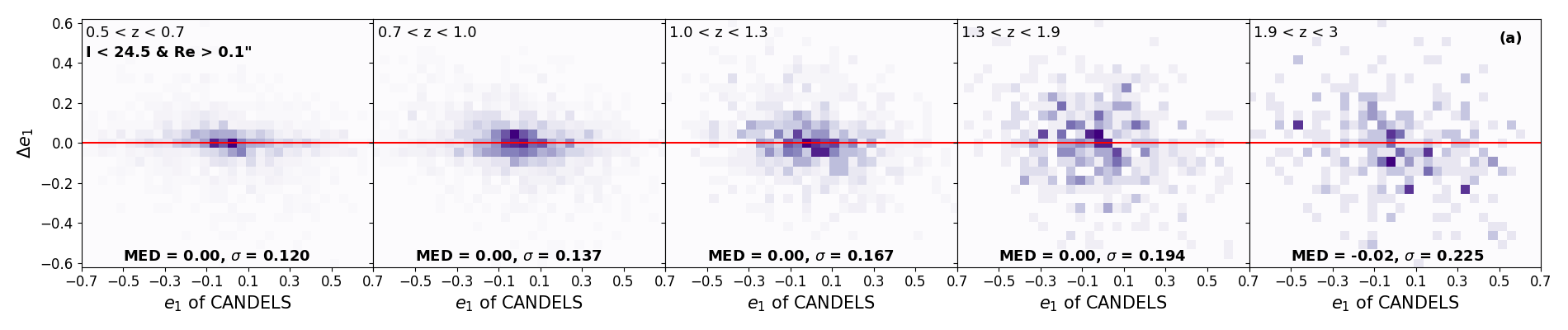}
\plotone{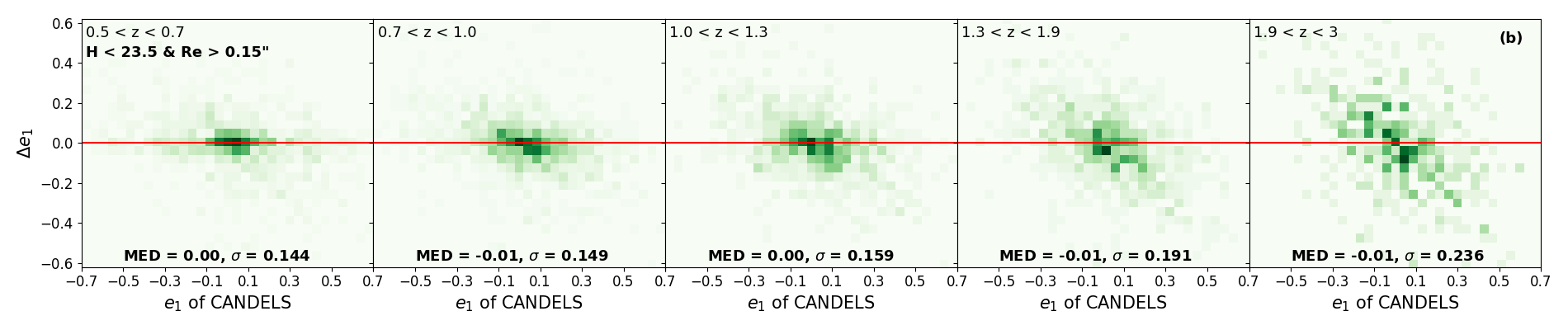}
\plotone{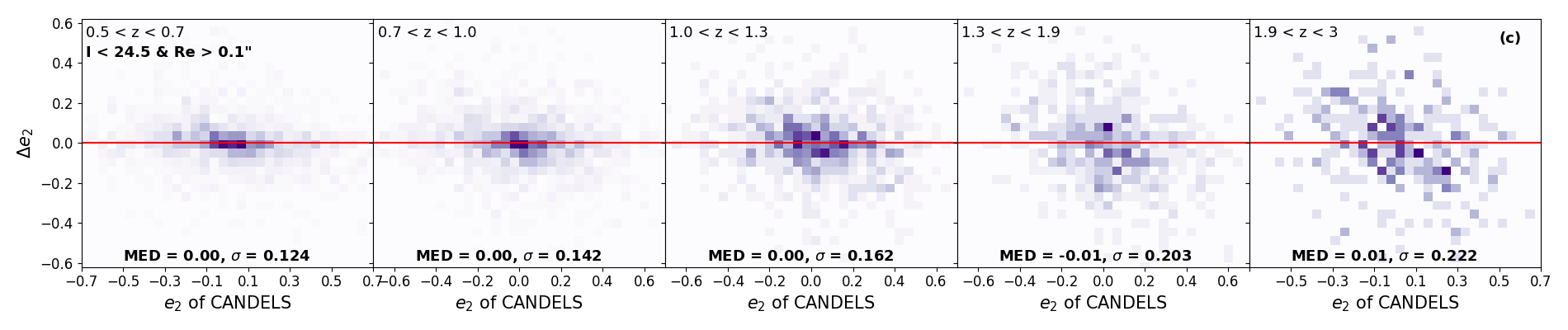}
\plotone{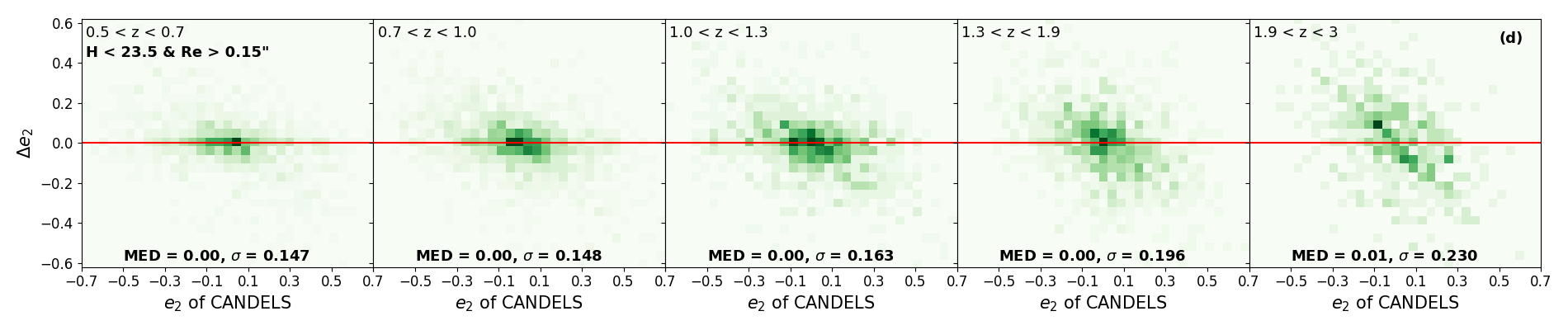}
\caption{We show the comparisons of the complex galaxy ellipticities derived using {\it Euclid}--quality and CANDELS images at five redshift bins. 
Here, we consider ellipticities from CANDELS single band image (I or H) as the true value and assess how well {\it Euclid} VIS and NIR-quality imaging can 
recover the original ellipticity of an observed galaxy. The purple and green plots represent the ellipticity comparisons for the 
optical (VI) and NIR selected samples (JH), respectively. The complex galaxy ellipticity 
consists of real ($e_{1}$) and imaginary parts ($e_{2}$) in the range of $-1<e_{1,2}<1$ as described by Equation~\ref{eq:ell}. 
The plot of $\Delta e_{1}$ = ( $e_{1}$ of Euclid-quality --  $e_{1}$ of CANDELS) 
vs. $e_{1}$ of CANDELS are shown in (a) and (b) for the optical and NIR samples. (c) and (d) show the plot of $\Delta  e_{2}$ vs. $e_{2}$ of CANDELS for 
the optical and NIR samples, respectively. The red line shows $\Delta  e_{1,2}=0$. Note that the darker 
color represents a denser region. The median and the standard deviation of $\Delta  e_{1, 2}$ for each redshift bin are written as 
MED and $\sigma$ on each plot. We find that most of galaxies are located around $\Delta e_{1,2} \sim 0$ in both the optical and NIR. 
The JH ellipticities are comparable in quality to the VI ellipticities, 
with the scatter in $\Delta e$ increasing with redshift due to the smaller sizes of galaxies and lower signal-to-noise ratio.}
\label{fig:complex}
\end{figure*}

\begin{figure*}
\centering
\epsscale{1.1}
\plotone{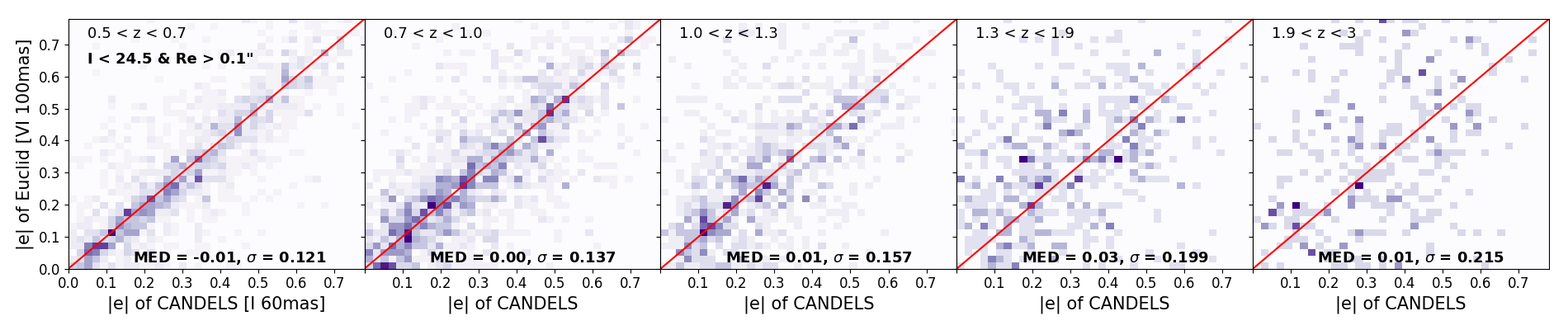}
\plotone{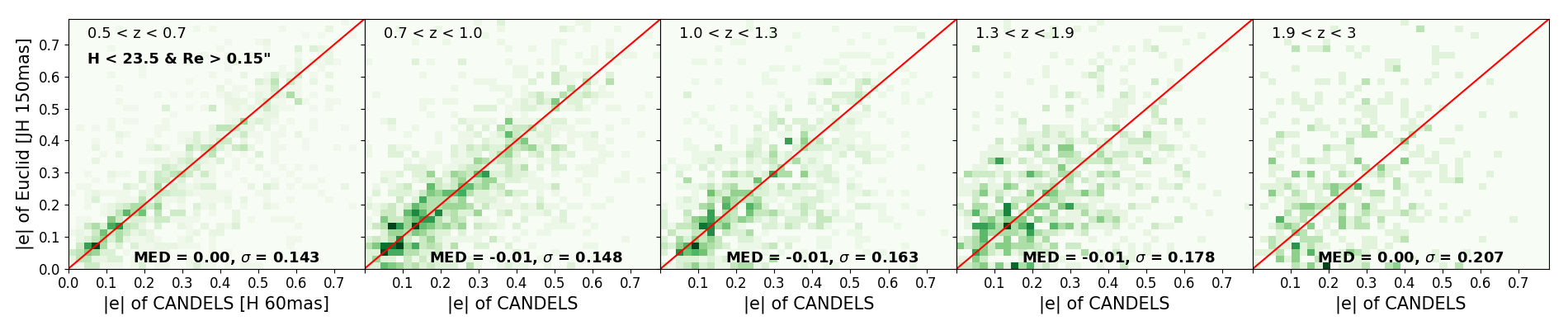}
\caption{Comparisons of the absolute values of galaxy ellipticities ($|e|$) derived using {\it Euclid}--quality and CANDELS images at five redshift bins. 
As in Figure~\ref{fig:complex}, the purple and green represent the comparisons for the optical and NIR samples, respectively. Note that the darker 
color represents a denser region. Red lines are one-to-one correlation. Top purple panels: 
$|e|$ derived using {\it Euclid}--quality VI images with 0.1$\arcsec$ pixel-scale plotted against $|e|$ derived using CANDELS I--band images with 0.06$\arcsec$. 
Bottom green panels: $|e|$ derived using {\it Euclid}--quality JH images with 0.15$\arcsec$ versus $|e|$ derived using CANDELS H--band images with 0.06$\arcsec$. 
We compute the median and standard deviation of 
$\Delta |e|$ = ( $|e|$ of Euclid-quality -- $|e|$ of CANDELS) and show them as MED and $\sigma$ on each plot. The ellipticities 
derived from Euclid-quality images show a very good correlation with the true values from CANDELS with MED$\sim 0$ at all redshift ranges. 
Overall, the scatter in JH-derived ellipticities is similar to the one measured in the VI-band. }
\label{fig:ell}
\end{figure*}

In order to study galaxy shapes, the complex galaxy ellipticity is typically used in weak lensing studies \citep{mil13, sch15}. Using axis 
ratio (q) and position angle estimated from GALFIT, we compute the complex galaxy ellipticity ($e$) of our final sample which is defined as
\begin{equation}
e = e_{1} + \mathrm{i}e_{2} = |e|\mathrm{e}^{2\mathrm{i}\phi}, 
\label{eq:ell}
\end{equation}
where, the modulus of the ellipticity ($|e|$) is defined as (1-q)/(1+q) and 
$\phi$ corresponds to the position angle of the major axis.
We then compare complex galaxy ellipticities ($e$) measured from Euclid-quality images with the CANDELS values. 
Here, we consider CANDELS-measured ellipticity as the original ellipticity of an observed galaxy because of 
the much larger depth and better resolution compared to {\it Euclid} data. Through this comparison, we can investigate the robustness of the galaxy shape measurements 
on the simulated {\it Euclid} images. 

In Figure~\ref{fig:complex}, we plot 
differences in ellipticities between CANDELS and {\it Euclid}-quality data, $\Delta e_{\alpha}$ = ($e_{\alpha}$ of Euclid-quality --  $e_{\alpha}$ of CANDELS), 
as a function of $e_{\alpha}$ of CANDELS for both ellipticity components, $\alpha=1,2$, in the range of $-1<e_{\alpha}<1$. 
Most galaxies scatter systematically around $\Delta e_{\alpha} = 0$ for both optical (purple) and NIR (green) with a median $\sim 0$ at all 
redshift ranges considered in this study. The uncertainty in $\Delta e$ quantifies how well the galaxy shapes are recovered with Euclid-quality images. 
The lowest redshift bin has a measured scatter ($\sigma$, the standard deviation of $\Delta e_{\alpha}$) in the NIR which is a factor of 1.2 larger than that in the VI-band.
However, the measured scatter  in the NIR is similar to that in the VI at all other redshifts, $z>0.7$.
Overall, we find that 
the ellipticities of individual galaxies can be measured with a similar scatter from the Euclid VIS- and NIR-like images.
There is a weak trend that the scatter of $\Delta e_{\alpha} $ increases with redshifts in both selections. Galaxies are fainter and smaller at higher redshift;
so, the limited {\it Euclid} spatial resolution and sensitivity will result in a larger scatter in the measured shape.
In particular, at higher redshifts, galaxies with larger ellipticities (in absolute 
values) tend to have larger discrepancies (see diagonal trends at $z>1.3$). This is likely because highly elongated galaxies in CANDELS appear to be rounder, and
with less-constrained position angles
at {\it Euclid}-quality resolution.
This trend appears to be a bit stronger for the NIR 
high-z sample due to the pixelization in the JH data.

\begin{figure}
\centering
\epsscale{1.1}
\plottwo{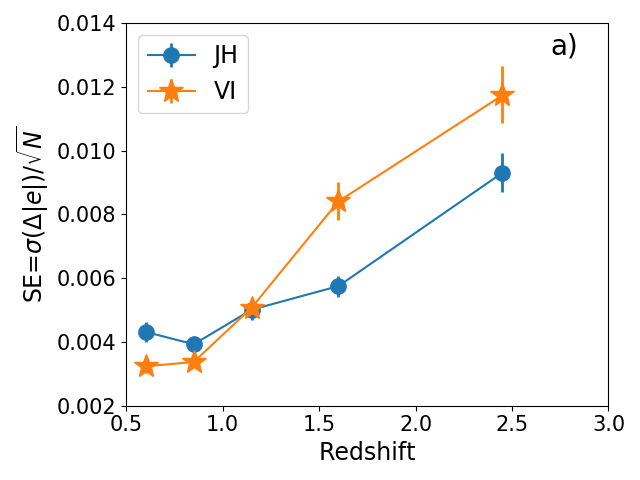}{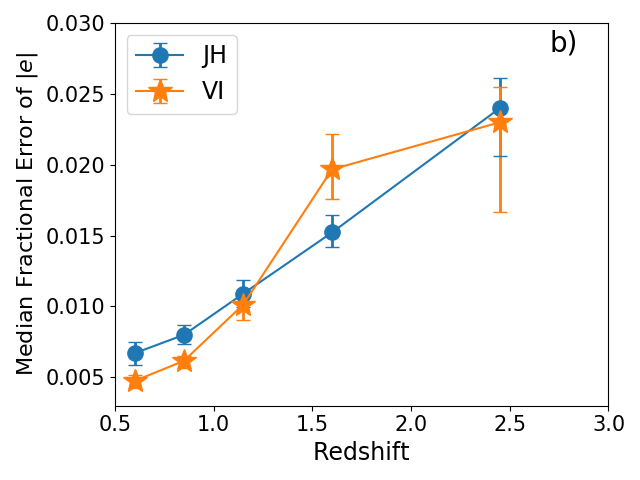}
\caption{a) Uncertainty of the ellipticity difference between CANDELS and simulated {\it Euclid}-quality images (i.e. standard error (SE) $\equiv \sigma(\Delta |e|)/\sqrt{N}$) as a function of redshift with error-bars determined by bootstrapping the sample. 
The SE in JH is a factor of 1.5 and 1.3 smaller than VI at $z\sim1.6$ and $z\sim2.5$.  b) Median fractional error of ellipticity (= Median($\frac{\Delta |\it{e}|}{|\it{e}| [CANDELS]})/\sqrt{N}$) as a function of redshift with error-bars determined by 
bootstrapping the sample. We find that VI and JH imaging yields ellipticities that are consistent with each other at all redshifts.
Note that N is the number of galaxies in each redshift bin over 0.1 $deg^{2}$.}
\label{fig:sigma}
\end{figure}

In Figure~\ref{fig:ell}, we compare the modulus of the ellipticity, $|e|$ from Equation~\ref{eq:ell}, derived using CANDELS I band to the simulated {\it Euclid} VI
and CANDELS H band to the simulated {\it Euclid} JH. The ellipticities derived from the simulated {\it Euclid} images are 
correlated very well with the ellipticities derived from CANDELS images with a median of $\Delta |e|$ = ($|e|$ of Euclid-quality -- $|e|$ of CANDELS) $\sim 0$ 
for both optical and NIR imaging. At $z>1.0$, NIR yields a similar to lower scatter than the 
optical, while the scatter in VI-derived ellipticities is significantly smaller at $z<0.7$. This  trend of the uncertainty in $\Delta \it{e}$ is more obvious in Figure~\ref{fig:sigma}--a). 
We compare the standard error (SE) of $\Delta |e|$ ($=\sigma(\Delta |e| / \sqrt{N}$) at each redshift bin for JH (blue) and VI (red). At $z<1.0$, 
the standard error of JH is 1.2--1.1 times larger than one of VI. But, the trend reverses at $z>1.0$ so that the standard error of JH is significantly smaller than VI by a factor of 1.5--1.3. 
As shown in Figure~\ref{fig:ell}, the measured $\sigma(\Delta |e|)$ of galaxies in JH is very similar with that of VI over the redshift range considered here; thus, the higher number 
density of galaxies at $z>1.0$ drives a lower standard error of ellipticity differences in JH. In Figure~\ref{fig:sigma}--b, we show the median fractional error of ellipticity, which is 
defined as (median of $\frac{\Delta |e|}{|e|[CANDELS]}$)$/\sqrt{N}$, as a function of redshift. The small values indicate that the derived ellipticity with $Euclid$-quality JH and VI imaging is very 
close to the true value from CANDELS on average. We find that the performance of the JH band in galaxy ellipticity measurements is comparable to the VI at all redshifts despite the 
significantly worse spatial resolution of JH. 
This result is very similar to that derived by \citet{tun17} who found that for a 1.2m class telescope, the $K_{s}-$band yields an ellipticity measurement 
error which is a factor of $\sim$3 smaller than in an $R-$band selected catalog while the $J-$band is a factor of $1.5-2.5$ worse than the $K_{s}-$selected catalog. 
This is also consistent with the results of \citet{sch18} who found that ground-based $K_{s}$ imaging with a PSF FWHM$\sim$0.35$\arcsec$ yields an ellipticity
 dispersion for $z\gtrapprox$ 1.4 galaxies, which is 0.76 times that of optically-selected galaxy samples with single-orbit {\it HST} imaging. 
A comparison between ellipticities derived from the simulated {\it Euclid} JH data and CANDELS I-band data indicates a correlation; however, the I-band ellipticities are 
larger than that in the NIR and the scatter is larger than shown in Figure~\ref{fig:ell}. This is likely because the $I-$band ellipticities are dominated by disk light while the NIR 
ellipticities are tracing a combination of disk and more-compact bulge light. Thus, if systematics arising from PSF under-sampling and dither strategy on the NIR images 
can be accounted for in future work, the shape noise can be minimized by including ellipticity measurements from the NIR bands.

\section{Conclusion: Precise Ellipticity Measurements in the NIR}

We investigate galaxy ellipticities in simulated 
{\it Euclid}-quality optical (VIS) and near-infrared (NIR) images constructed from {\it HST}/CANDELS co-added V+I (VI) and J+H (JH) images. We select galaxies in CANDELS 
GOODS-S and -N fields (covering about 340 arcmin$^2$) with photometry in I and H bands at similar depths as the planned 
{\it Euclid} survey. In this study, we specifically use galaxies at a redshift range of $0.5\le z<3$. 
After applying a SNR $\gtrapprox 10$ cut,
the total number density of galaxies at $0.5\le z<3$ is comparable in the NIR and optical; however, the NIR bands select 1.4 times higher number density of galaxies 
relative to the optical selection at $z>1.0$, which enable us to reduce the statistical uncertainties in the shape measurements of distant galaxies.
By co-adding {\it Euclid}-quality J and H-band images, which double the number 
of frames and the exposure time, we can generate
images at 0.15$\arcsec$ pixel scale (half of the original {\it Euclid} NIR detector pixel scale) and 
achieve $S/N \gtrapprox 7$ at $H<24$\,mag and $\gtrapprox11$ at $H<23.5$\,mag. 
Using GALFIT, we measure ellipticities of galaxies in CANDELS I and H band images with 0.06$\arcsec$ pixel scale and 
the simulated {\it Euclid} VI and JH images with 0.1$\arcsec$ and 0.15$\arcsec$ pixel scale, respectively. 
We then compare ellipticities between CANDELS and simulated {\it Euclid}-quality images while considering {\it HST}/CANDELS ellipticity as the 
original ellipticity of a galaxy due to its superior depth and resolution. 
 
A comparison between ellipticities derived from CANDELS and {\it Euclid}-quality VIS and NIR 
imaging shows that both wavelength ranges provide similar performance in measuring galaxy ellipticities at all redshifts included in this study despite
the worse spatial resolution and pixel sampling of the NIR imaging. When combined with the
higher source density in the NIR selection, we find that the standard error in NIR-derived ellipticities is about 30\% smaller than the optical bands at $z>1.0$, which 
implies a more precise ellipticity measurement than in the optical alone. Since the VIS and NIR galaxy shape measurements with {\it Euclid} have different fractional contributions
of the bulge and disk, a combination of the two
can improve the precision with which galaxy ellipticities are measured. 
The next step that is required before the NIR data can be used for WL studies is to assess how the drizzling affects both the 
undersampled telescope PSF, and the correlated noise \citep[see e.g.][]{rho07}.
However, even though the FWHM of the {\it Euclid} NIR imaging does not quite reach {\it HST} or WFIRST resolution, the NIR data provides a 
major advantage for weak lensing measurements compared to optical ground-based observations that typically achieve a PSF FWHM$\sim 0.6\arcsec-0.7\arcsec$ in 
good seeing conditions \citep[e.g.][]{kui15, man18}; while the latter
provides good sensitivity to the weak lensing signal with a median 
redshift of the sample of z $\sim 0.85$, about half the galaxy sample will be unresolved due to the 
small size of galaxies, as shown in Figure~\ref{fig:re}, implying a higher statistical uncertainty in their ellipticities. 

In conclusion, by using co-added J+H band {\it Euclid}--quality images, we show that the galaxy sample selected at NIR wavelengths yields a more precise ellipticity measurement, 
especially at high redshifts. This suggests that a careful evaluation of 
NIR shape systematics for future weak gravitational lensing surveys, such as with {\it Euclid} and WFIRST, should be undertaken.

\acknowledgements This work is partly funded by NASA/{\it Euclid} grant 1484822 and is
based on observations taken by the CANDELS Multi-Cycle Treasury Program with the NASA/ESA HST, which is operated by the Association of Universities for Research in Astronomy, Inc., under NASA contract NAS5-26555. The authors thank the anonymous referee for very useful comments that helped to improve the presentation of the paper. We also thank Lance Miller, Stefanie Wachter, Peter Schneider, Henk Hoekstra and Jason Rhodes for thoughtful comments which improved this manuscript.


\begin{thebibliography}{}
\bibitem[Bartelmann \& Schneider(2001)]{bar01}
Bartelmann,M. \& Schneider, P. 2001, Phys. Rep., 340, 291
\bibitem[Bertin \& Arnouts(1996)]{ber96}
Bertin, E. \& Arnouts, S. 1996, A\&AS, 117, 393
\bibitem[Dahlen et al.(2013)]{dah13}
Dahlen, T., Bahram, M., Faber, S. M., Ferguson, H.C., et al. 2013, \apj, 775, 93
\bibitem[Dickinson(2000)]{dic00}
Dickinson, M. 2000, Philos. Trans. R. Soc. London A, 358, 2001
\bibitem[Grogin et al.(2011)]{gro11}
Grogin, N. A., Kocevski, D. D., Faber,S. M. et al. 2011, \apjs, 197, 35
\bibitem[Cropper et al.(2012)]{cro12}
Cropper, M., Cole, R., James, A., et al. 2012, Arxiv:1208.3369
\bibitem[Hoekstra et al.(1998)]{hoe98}
Hoekstra, H., Franx, M., Kuijken, K., \& Squires, G. 1998, \apj, 504, 636
\bibitem[Holden et al.(2009)]{hol09}
Holden, B. P., Franx, M., Illingworth, G. D., et al. 2009, \apj, 693, 1
\bibitem[Jee et al.(2013)]{Jee} Jee, M.~J., Tyson, J.~A., Schneider, M.~D., et al.\ 2013, \apj, 765, 74 
\bibitem[Kaiser et al.(1995)]{kai95}
Kaiser, N., Squires, G., \& Broadhurst, T. 1995, \apj, 449, 460
\bibitem[Koekemoer et al.(2011)]{koe11}
Koekemoer, A. M., Faber, S. M., Ferguson, H. C. et al. 2011, \apjs, 197, 36
\bibitem[Kitching et al.(2008)]{kit08}
Kitching, T. D., Miller, L. Heymans, C. E. et al. 2008, \mnras, 390, 149
\bibitem[Kuijken et al.(2015)]{kui15}
Kuijken, K., Heymans, C., Hildebrandt, H., et al. 2015, \mnras, 454, 3500
\bibitem[Krist(1995)]{kri95}
Krist, J. 1995, Astronomical Data Analysis Software and Systems IV, 77, 349
\bibitem[Laureijs et al.(2011)]{lau11}
Laureijs, R., Amiaux, J., Arduini, S. et al. 2011, arXiv: 1110.3193
\bibitem[Mandelbaum et al.(2018)]{man18}
Mandelbaum, R., Miyatake, H., Hamana, T. et al. 2018, \pasj, 70, S25
\bibitem[Massey et al.(2013)]{mas13}
Massey, R., Hoekstra, H., Kitching, T., et al. 2013, \mnras, 429, 661
\bibitem[Miller et al.(2007)]{mil07}
Miller, L., Kitching, T. D., Heymans, C., et al.  2007, \mnras, 382, 315
\bibitem[Miller et al.(2013)]{mil13}
Miller, L., Heymans, C., Kitching, T. D., et al. 2013, \mnras, 429, 2858
\bibitem[Peng et al.(2002)]{pen02}
Peng, C. Y., Ho, L. C., Impey, C. D., \& Rix, H.-W. 2002, \aj, 124, 266
\bibitem[Rhodes et al.(2007)]{rho07} 
Rhodes, J.~D., Massey, R.~J., Albert, J., et al.\ 2007, \apjs, 172, 203 
\bibitem[Schrabback et al.(2015)]{sch15}
Schrabback, T., Hilbert, S., Hoekstra, H., et al. 2015, \mnras, 454, 1432
\bibitem[Schrabback et al.(2018)]{sch18}
Schrabback, T., Schirmer, M., van der Burg, R. F. J. et al. 2018, \aap, 610, A85
\bibitem[Sif\'{o}n et al.(2015)]{sif15}
Sif\'{o}n, C., Hoekstra, H., Cacciato, M. et al. 2015, \aap, 575, 48
\bibitem[Troxel \& Ishak(2015)]{tro15}
Troxel, M., A., Ishak, M. 2015, Phys. Rep., 558, 1
\bibitem[Tung \& Wright(2017)]{tun17} 
Tung, N. \& Wright, E.\ 2017, \pasp, 129, 114501 
\bibitem[van der Wel et al.(2012)]{vdw12}
{van der Wel}, A. and {Bell}, E.~F. and {H{\"a}ussler}, B. et al. 2012, \apjs, 230, 24
\end{thebibliography}
\end{document}